# Non-linear double-peeling: experimental vs. theoretical predictions

D. Misseroni[(1)], L. Afferrante[(2)], G. Carbone[(2)], N. Pugno[(1, 3, 4, 5*)]

[(1)] Instabilities Lab, Department of Civil, Environmental and Mechanical Engineering, University of Trento, Trento, Italy;

[(2)] TriboLAB – Laboratory of Tribology, Department of Mechanics, Mathematics and Management (DMMM) Politecnico of Bari, v.le Japigia 182, 70126 Bari, Italy.

[(3)] Laboratory of Bio-inspired & Graphene Nanomechanics, Department of Civil, Environmental and Mechanical Engineering, University of Trento, Trento, Italy;

[(4)] School of Engineering and Materials Science, Queen Mary University, Mile End Rd, London E1 4NS, UK.

[(5)] Italian Space Agency, Via del Politecnico snc, 00133 Rome, Italy

[*]Corresponding author: nicola.pugno@unitn.it, **Tel:** +39 0461 282525, **Fax:** +39 0461 282599

## Abstract

The double peeling of detachment of non-linear adhesive tapes from a flat Poly(methyl methacrylate) (PMMA) surface has been investigated from both experimental and theoretical point of view. Double peeling tests show that, as the detachment process advances, the peeling angle stabilizes on a limiting value $\theta_{\text{lim}}$ corresponding to a critical pull-off force $F_c$ above which the tape is completely detached from the substrate. This observed behavior is in good agreement with results obtained following the new theory of multiple peeling and taking into account the hardening-softening non-linear behavior of the experimentally tested adhesive tapes and clarifies some aspects of the experimental data. In particular, the theoretical model shows that the value of the limiting peeling angle depends on the geometry of the adhesive tape as well as on the stiffness properties and on the interfacial energy $\Delta\gamma$. Finally, theoretical predictions confirm that solutions with a peeling angle lower than $\theta_{\text{lim}}$ are unstable.

Keywords: double-peeling, adhesion of tapes, energy release rate, thin film peeling, van der Walls adhesion.

# 1 Introduction

Biomimetic microstructured surfaces show enhanced adhesive [1-3] or superhydrophobic [4-7] properties and, for this reason, they have been attracting more and more scientific interest. Starting with research [8-19] on the adhesion behavior of lizards (like gecko, for instance) and arthropods (like beetles or spiders), a great deal of experimental [20-24] and theoretical [25-31] work has been done to study the mechanism of adhesion and detachment of natural systems.

The angle giving optimal adhesion could be used in nature by animals to maximize adhesion at all the different hierarchical levels, such as contralateral legs, toes and even setae for geckos [23]. In addition, the optimal angle could also be used in nature to optimize the strength of natural anchors [8, 26]. For example, in order to maintain the necessary shear/frictional adhesive forces and to avoid toe detachment, the Tokay gecko's (*Gekko gecko*) adhesive mechanism is based on the use of opposing feet and toes making a V-shaped geometry. The gecko then pulls its feet inwards towards the center of its body mass and its toes inwards towards the foot to engage adhesion [23, 10].

The key factor that governs the gecko's attachment/detachment mechanism is the adhesion angle $\alpha$ between the terminal structure attached to the surface and the surface itself. Several scientific studies have been developed to establish the value of such an adhesion angle $\alpha$ from an experimental [9-10, 21], theoretical or computational [12-13, 18, 32] point of view and at different characteristic sizes of the hierarchical adhesive system. From previously published scientific papers, the adhesion angle $\theta$ of Tokay geckos is equal to ~29° [23] for opposing front and rear feet, ~27.3 [23] between the first and fifth toe of each foot, ~25.5° [20] for a single toe, ~24.6° [20] (or ~30° [22]) for isolated setae arrays and ~30.0° [20] (or ~31° [22]) for a single seta.

Beside the pioneering work of Kendall [33], there exist numerous other theoretical [34-39] and experimental [40-43] studies on tape peeling and recently its extension to multiple peeling has also been deduced [26]. In particular, the influence on the peeling force of the bending stiffness has been investigated in Ref. [39], by finite element computations. In Ref. [40], instead, the peeling dynamics of an adhesive tape has been studied with special attention on the stick-slip regime of the peeling. The effect of pre-tension on the critical detachment angle in gecko adhesion has been also addressed in Refs. [44-45], where it has been found that beyond a critical pre-tension the peel-off force vanishes at a force-independent critical peeling angle.

In this paper, we experimentally estimated the adhesion angle and the peeling force of a standard adhesive tape through double peeling tests in order to verify the existence of an optimal adhesion angle. In general, the findings presented here could be useful for the industrial fabrication of bio-inspired dry adhesives tapes, robotics systems, artificial adhesive suits and gloves for astronauts or to design bio-inspired adhesive nanomaterials and adhesive biomaterials, such as smart biomedical patches or self-adhering bandages and dressings.

Moreover, the proposed geometry could be of interest in many natural systems (most biological hairy adhesive systems involved in locomotion rely on spatula-shaped terminal elements [28], the spider silk anchors are often V-shaped and hierarchically arranged [26]). For such type of systems, the multiple peeling model [26] could be employed to explain the reason for appearance of more tape-like terminal elements at the smallest hierarchical level.

## 2 Double peeling of adhesive tapes

### 2.1 Experimental Setup

All the experiments were performed using a Tesa® Universal tape (Tesa SE, Germany) with a mean thickness of ~90 μm and a width of 15 mm.

The mechanical characteristics of the adhesive tape were evaluated by means of tensile tests, performed on 5 specimens, using a Midi 10 electromechanical testing machine (10 kN maximum force, from Messphysik Materials Testing, Fürstenfeld, Austria). The Midi 10 was used to impose a constant displacement (speeds of 0.1 mm/s) at the ends of the tape. Load and displacement were measured with the load cell MT 1041 (R.C. 10 kg) and the displacement transducer mounted on the Midi 10 machine. Data from both the load cell MT 1041 and the displacement transducer was acquired using NI CompactRIO system interfaced with Labview 2013 (National Instruments).

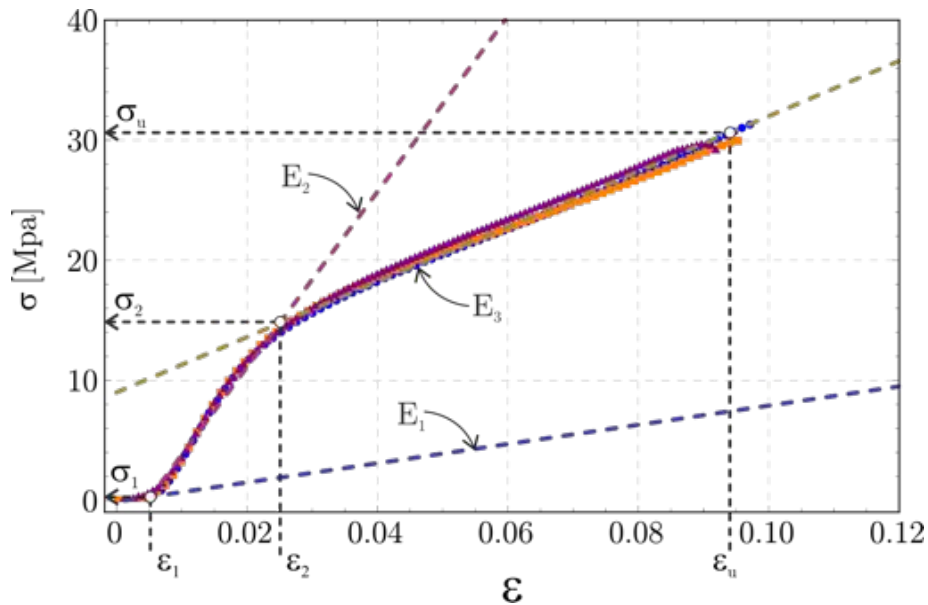

**Fig. 1.** Tensile test of some samples, as representative of the general trend of all tested samples. The piecewise linear interpolation used in the following calculations is also shown.

During tensile tests, specimens were pulled until they broke completely. Then, the stress-strain curves of tensile tests were computed using the estimation of the real width and thickness at the cross-section of each specimen.

Figure 1 shows the stress-strain curve obtained from the tensile test on some samples, as representative of the general trend of all tested samples. We notice the experimental curve shows a hardening-softening non-linear behavior. However, a piecewise linear model can be implemented in order to fit the constitutive relation. Indeed, the stress-strain curve is well fitted by three linear segments, with different slope. The elastic modulus $E$ of each segment was calculated as the slope of the interpolating line. The area under the stress-strain curve supplies the energy required to break the material and corresponds to the toughness modulus $U_t$. The results obtained for ultimate stress, ultimate strain, elastic moduli and toughness are shown in Table 1.

| $\sigma_u$ [MPa] | $\varepsilon_u$ | $\varepsilon_1$ | $\varepsilon_2$ | $E_1$ [MPa] | $E_2$ [MPa] | $E_3$ [MPa] | $U_t$ [MJ/m$^3$] |
|---|---|---|---|---|---|---|---|
| 30.8 | 0.094 | 0.0052 | 0.025 | 80 | 730 | 230 | 1.72 |

Table 1. Results of tensile tests: mean values of the mechanical properties of the adhesive tape.

In order to evaluate the work of adhesion per unit area $\Delta\gamma$, also referred to as the Dupré energy of adhesion [46], five 180°-peeling tests of the same tape were performed on the PMMA surface, in line with UNI EN 1939 for the determination of the peel adhesion properties. The experimental setup, specifically designed and manufactured for performing these tests, is reported in Fig. 2a. The tests were performed imposing displacement (speeds of 0.1 mm/s) at one end of the tape. The load were measured by a load cell with a rated capacity of 10 N.

The work of adhesion $\Delta\gamma$ has been obtained by the Griffith's criterion, using the measured values of the force per unit length required to remove the tape. We have shown the comparison between experimental results and theoretical predictions, using for $\Delta\gamma$ the average measured value of (74 J/m$^2$) as obtained from the 180°-peeling tests. We anticipate that the proposed theoretical model does not consider the potential viscous behaviour that can occur during the peeling process. For this reason, we have preferred to use an 'effective' value of the work of adhesion as a measure of the tape resistance to detachment. Such effective value, obtained on tests performed in similar conditions, is then used to compare the theoretical predictions with the experimental results.

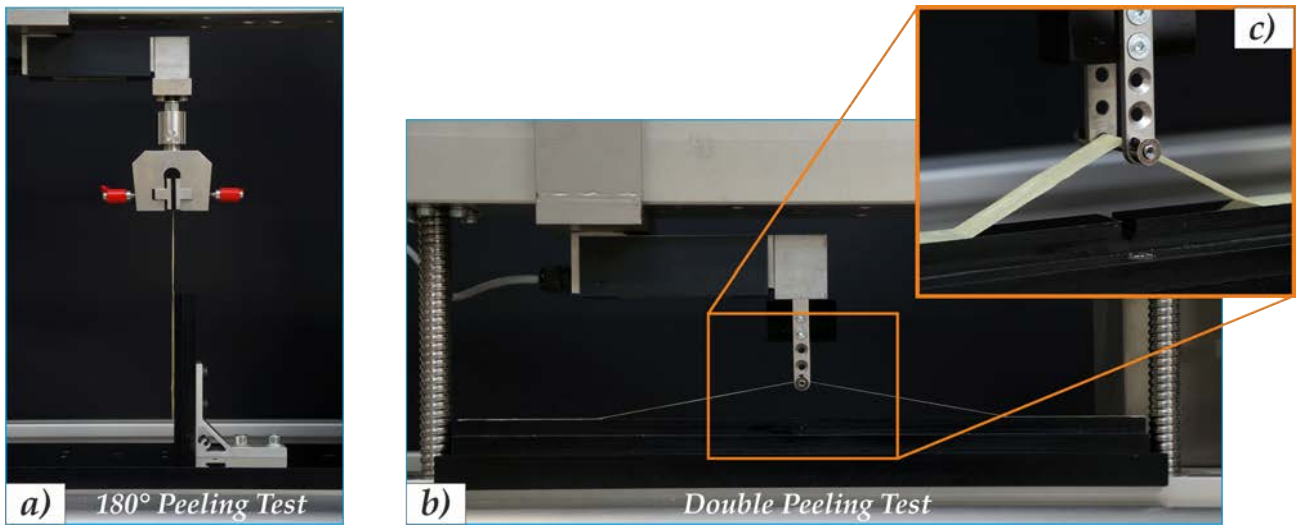

**Fig. 2.** The experimental setup employed for the 180° peeling test a) and for the double peeling test b). c) Detail of the specifically designed loading device that provide the symmetric loading condition at the middle of the tape

.

Double peeling tests were performed on 10 specimens of the same adhesive tape, using a the Midi 10 testing machine equipped with a 1 kg load cell in order to estimate the force ($F_c$) to detach the tape at a speed of 0.1 mm/min. A specific experimental setup (Fig. 2b) were designed and realized at the Laboratory of the University of Trento, to ensure that the experiments respect the assumptions at the base of the theoretical model. a An ad hoc support for the tape (width of 15mm) was realized by milling a PMMA sheet with a EGX-600 Engraving Machine (by Roland).. This system ensure that the tape is perfectly aligned and that the force is applied exactly in the middle and perpendicular to the adhesive tape as is clearly visible in Fig. 2b. The force in the middle was provided by the specific equipment showed in Fig. 2c that provide the symmetric loading condition at the middle of the tape. In the middle of the PMMA support was realized a seat for the linchpin that provide the load to the tape in order to ensure that at the beginning of all the tests the tape was perfectly stuck on the PMMA support without induce any undesired pre-stress within the tape.

A specific experimental protocol was followed to ensure the repeatability of the tests. First of all the PMMA support was cleaned, before each test, with an ethanol soaked cloth. The adhesive tape was then stuck to the support imposing a constant force along all the length of the tape by a rigid pad In so doing the adhesion along the tape is as much uniform is possible. The tests were performed starting with the adhesive tape completely stuck to the PMMA base. Finally, the load in the middle was imposed by means of a oiled linchpin connected to the moveable crosshead of the testing machine.

Finally, the tape specimens were double peeled until they completely detached. - The Labview program recorded the experimental data of the delamination force, which is applied to the tape and then the force-time curves were computed.

Meanwhile, tests were recorded by a HDR-XR550SONYdigital video camera, which was placed in front of the testing platform. The camera recorded a video from which we extracted 24 frame per second (Fig. 3). The frame sequence was elaborated by a Matlab, © 1994-2012 The MathWorks, Inc. code, which is specifically programmed for these specific experiments. This Matlab code is able to measure the two angles, $\theta_{sx}$ and $\theta_{dx}$, between the adhesive tape and the PMMA base by individuating the pixel contrast between the white or black pixels of the tape or the background, respectively. Note that the angle $\theta_{sx}$ corresponds to the angle on the left side of the screen and $\theta_{dx}$ corresponds to the angle on the right side. Angle measurements at each video frame were used to study the evolution of the peeling angle during the test. In this way, both the video recording and the double peeling tests have the same time scale so the corresponding angles and the delamination force could be easily matched together.

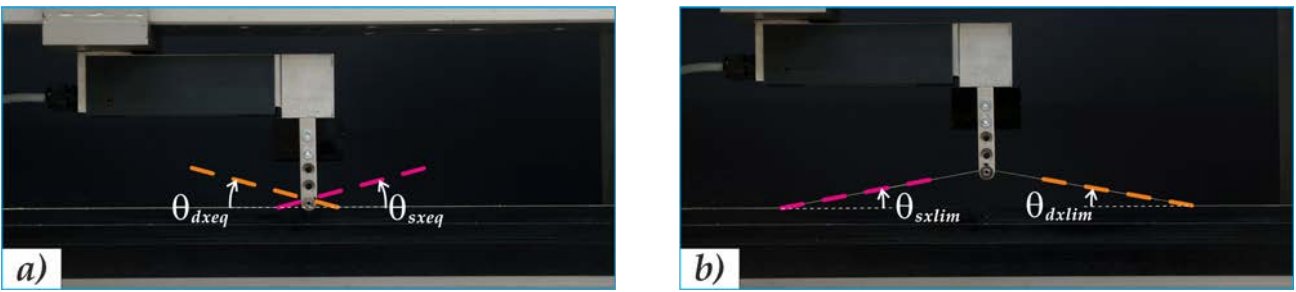

**Fig. 3**. Frames from a single test video at the beginning (a) and after few minutes (b) of a double peeling test (Ref. [31]).

## 2.2 Mathematical model

Fig. 4 shows a picture of an adhesive tape adhering to a smooth flat Poly(methyl methacrylate) (PMMA) surface and pulled apart with a vertical force $F = 2P$.

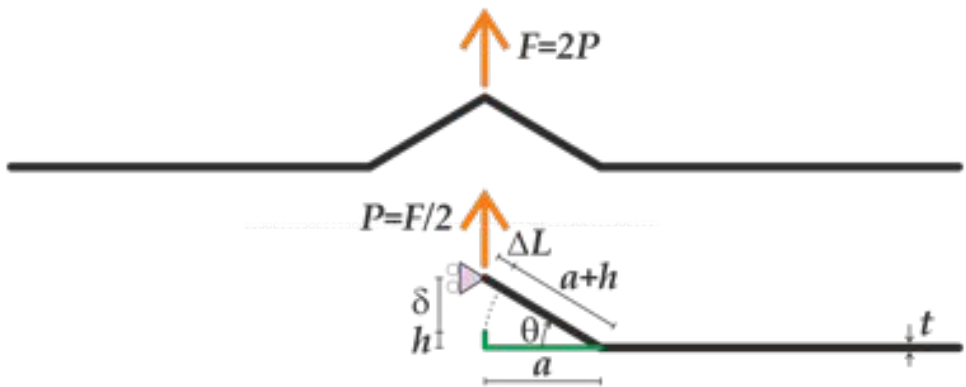

**Fig. 4**. Double-peeling of a tape

Due to the symmetry of the system, we can limit our study to half the tape. The action of the vertical force $P$ makes the tape increase its length of the quantity $\Delta L$

$$P,\theta = \varepsilon(P,\theta)(a+h) \tag{1}$$

where $a$ is the detached length, $h$ is a small quantity representing the local length of the tape that is not attached to the substrate (see Ref. [47]) and $\varepsilon(P,\theta)$ is the strain of the tape that, according to the stress-strain curve shown in Fig. 1, can be approximated by

$$\varepsilon(P,\theta) = \varepsilon_i + \frac{1}{E_{i+1}}\left(\frac{P}{bt\sin\theta} - \sigma_i\right) \qquad \sigma_i \leq \frac{P}{bt\sin\theta} \leq \sigma_{i+1}, \qquad i = 0, \ldots, n-1 \tag{2}$$

being $n$ the number of linear segments characterizing the constitutive relation ($n=3$ in our case), $\sigma_n$ the ultimate tensile strength and $\varepsilon_0 = \sigma_0 = 0$.

The elastic energy stored in the tape is then given by

$$U_{el} = \int_V \left(\int_0^\varepsilon \sigma(\varepsilon)d\varepsilon\right)dV = A(a+h)\sum_{i=0}^{n-1}\int_{\varepsilon_i}^{\varepsilon_{i+1}}[\sigma_i + E_{i+1}(\varepsilon - \varepsilon_i)]d\varepsilon \tag{3}$$

where $A$ is the cross sectional area of the tape, and the potential energy is

$$U_P = -P(a+h)[1 + \varepsilon(P,\theta)]\sin\theta \tag{4}$$

where the detached length $a = (a + h + \Delta L)\cos\theta$ can be written, in terms of the peeling angle $\theta$ and the vertical force $P$, as

$$\frac{a}{h} = \frac{[1+\varepsilon(P,\theta)]\cos\theta}{1-\cos\theta-\varepsilon(P,\theta)\cos\theta} \tag{5}$$

The energy release rate $G$ can be then obtained as

$$G = -\frac{1}{b}\left(\frac{\partial U_{el}(a,\theta(a))}{\partial a} + \frac{\partial U_P(a,\theta(a))}{\partial a}\right)_P = -\frac{1}{b}\left(\frac{\partial U_{el}}{\partial a} + \frac{\partial U_{el}}{\partial \theta}\frac{\partial \theta}{\partial a} + \frac{\partial U_P}{\partial a} + \frac{\partial U_P}{\partial \theta}\frac{\partial \theta}{\partial a}\right)_P \tag{6}$$

where the derivative $\partial\theta/\partial a$ is evaluated by eq. (5).

Since at equilibrium the total energy must be stationary, the energy release rate $G$, according to the Griffith criterion, must equal the work of adhesion $\Delta\gamma$. This equation allows relating the vertical pull-off force $P$ with the peeling angle $\theta$. Notice the present mathematical formulation is an extension to the non-linear case of the model developed in Ref. [31].

## 3 Results and discussion

In this section, we present results of the experimental tests and compare them with theoretical predictions.

### 3.1 Experimental results

The angles $\theta_{sx}$ and $\theta_{dx}$ and the corresponding delamination force $F_c$, were determined performing double peeling tests on 10 samples. The experimental setup employed in the tests is shown in Fig. 2b.

Fig. 5 shows an example of the evolution of the angles $\theta_{sx}$ and $\theta_{dx}$ as well as of the peeling force as a function of time. Notice both angles tend to stabilize around an almost constant value $\theta_{lim}$, corresponding to the final critical pull-off force above which complete detachment of the tape occurs.

Starting from a zero angle, the delamination starts after about 1 minute from the beginning of the test and causes a rapid variation of angles and forces.

Resulting data are given in Table 2, where the mean values of the limiting angles $\theta_{sx,lim}$, $\theta_{dx,lim}$ and the critical pull-off force $F_C = 2P_C$ are shown.

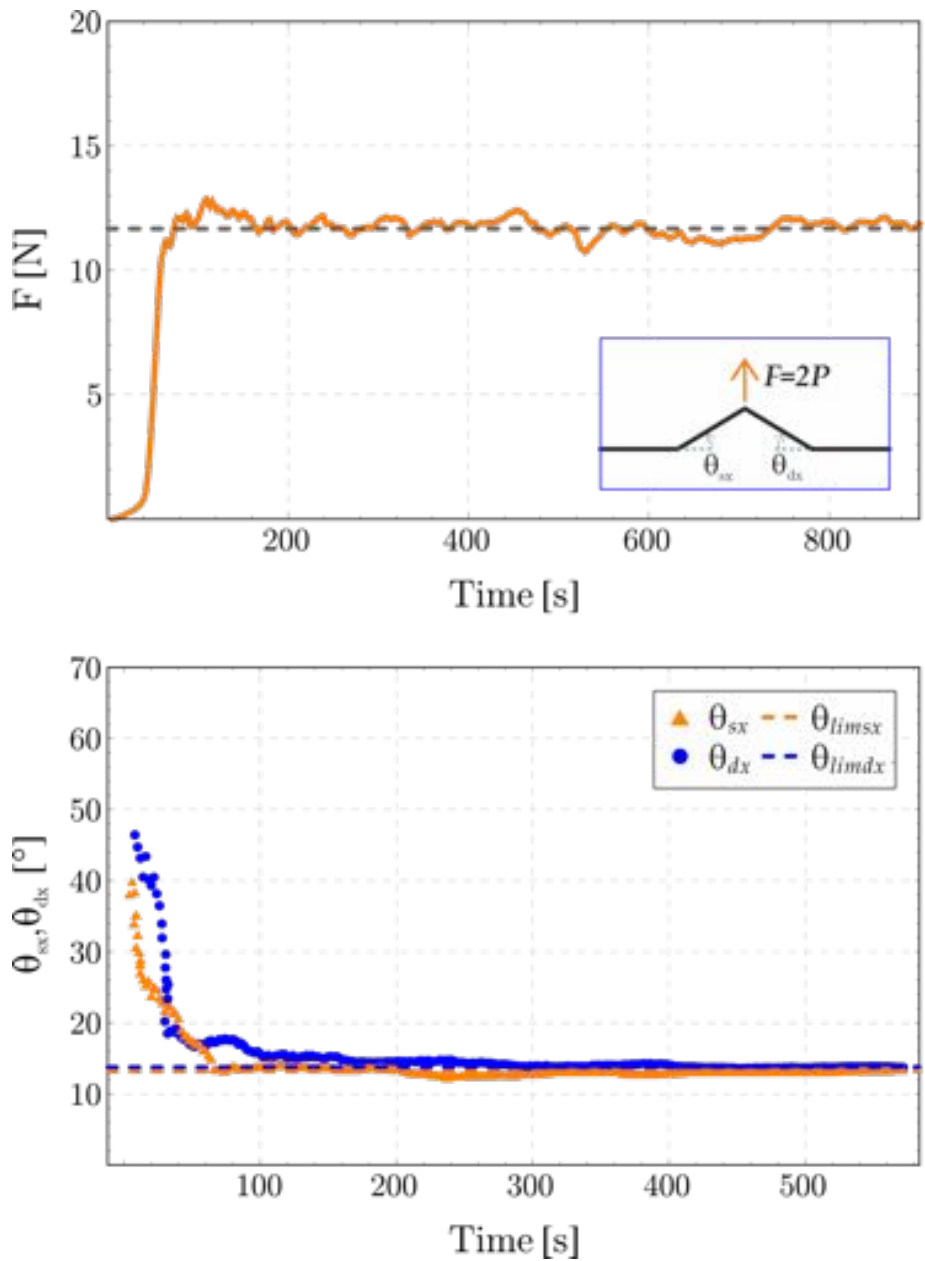

**Fig. 5.** Double peeling test of sample 5, as representative of the general trend of all tested samples. Variation of the peeling angle (a) and peeling force (b) with time.

| TEST | $\theta_{sx,lim}$ [°] | $\theta_{dx,lim}$ [°] | $F_c$ [N] |
|---|---|---|---|
| 1 | 13.21 | 13.75 | 11.90 |
| 2 | 14.04 | 13.70 | 11.04 |
| 3 | 14.19 | 13.35 | 10.46 |
| 4 | 14.57 | 13.25 | 11.12 |
| 5 | 15.59 | 14.47 | 12.68 |
| 6 | 13.77 | 13.56 | 11.25 |
| 7 | 14.62 | 13.97 | 11.67 |
| 8 | 13.90 | 14.61 | 11.12 |
| 9 | 14.32 | 14.40 | 10.98 |
| 10 | 14.68 | 13.88 | 11.91 |
| Mean value | 14.30 | 13.90 | 11.41 |
| St. Deviation | 0.636 | 0.468 | 0.62 |

**Table 2.** Experimental mean values for the limiting peeling angles $\theta_{sx,lim}$, $\theta_{dx,lim}$, the critical force $F_c$ and the corresponding stress in the tape.

### 3.2 Theoretical predictions and comparison with experimental data

Fig. 6 shows the pull-off force $F = 2P$ as a function of the peeling angle $\theta_{eq}$ at equilibrium, for the average measured values of the work of adhesion $\Delta\gamma$.

Notice by increasing the pull-off force the peeling angle decreases tending to a lower limit $\theta_{lim}$ (depending on $\Delta\gamma$), at which the pull-off force $F$ takes its maximum value $F_c$. Below $\theta_{lim}$ solutions are unstable (see the gray area in Fig. 6), as shown in [47], where it is demonstrated that the total energy takes a maximum value (and, hence, unstable) for peeling angles $\theta_{eq} < \theta_{lim}$.

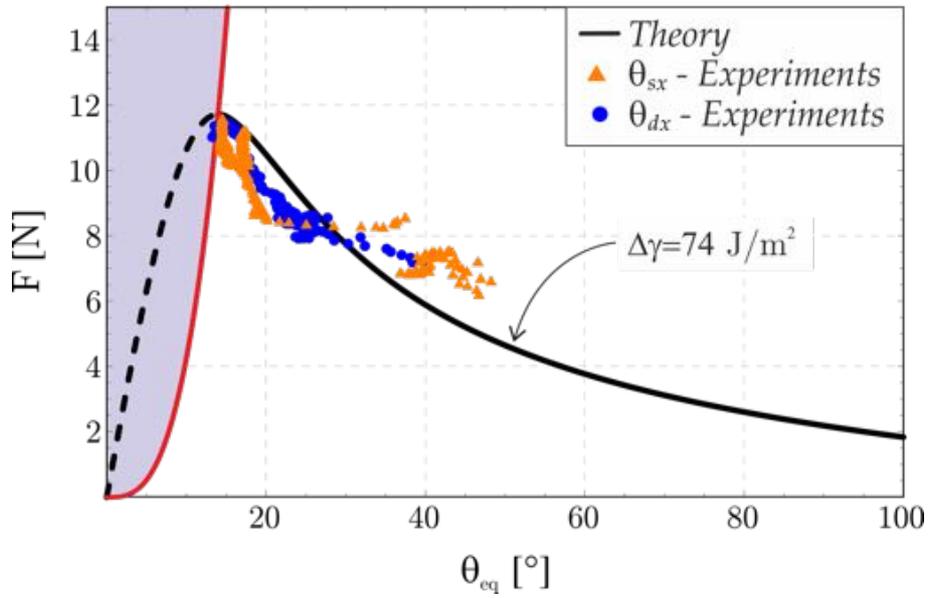

**Fig. 6**. The variation of the pull-off force with the peeling angle at equilibrium $\theta_{eq}$. Dashed lines denote the unstable branch of the curve. Experimental data refer to tests 5 and 6 as representative of the general trend of all tests.

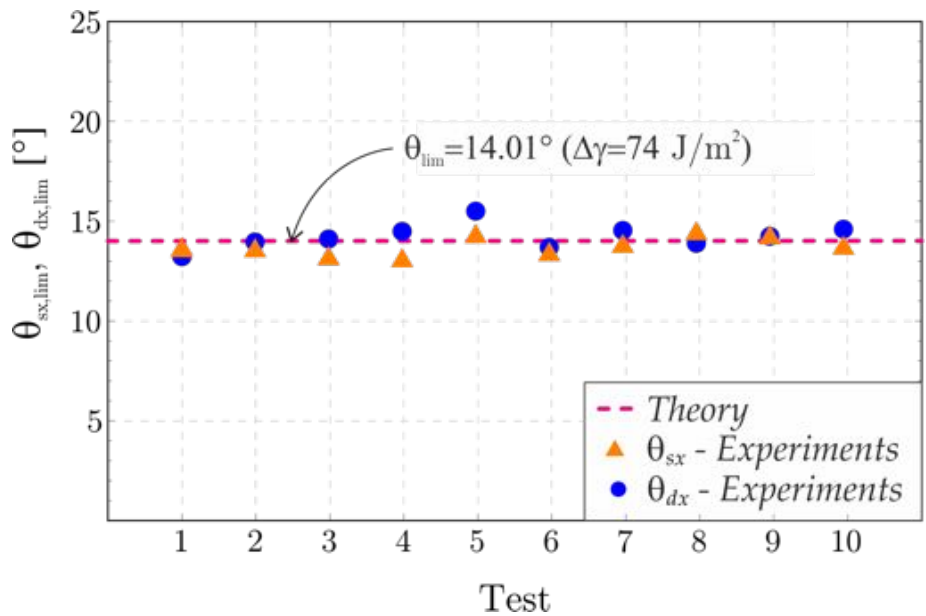

**Fig. 7**. Comparison between the experimental limiting peeling angles with the theoretical predictions (solid and empty circles refer to values of $\theta_{1,lim}$ and $\theta_{2,lim}$, respectively).

Experimental data are also plotted in the figure for comparison. The theoretical trend of the pull-off force properly fits experimental data at $\Delta\gamma$ = 74 J/m².

Figs. 7 and 8 show a comparison between experimental results and theoretical predictions in terms of limiting peeling angle and critical pull-off force $F_c$.

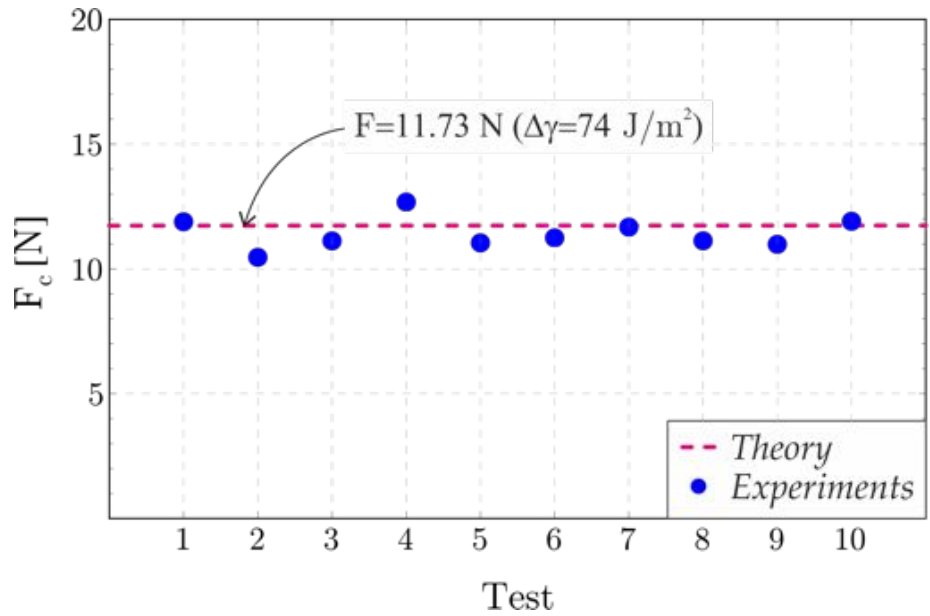

**Fig. 8**. Comparison between the experimental critical pull-off force with the theoretical predictions.

The measured limiting peeling angles are almost in high agreement with the angle defined by the theoretical values obtained for Δγ = 74 J/m² . Fig.

In our experiments, the tape is not very compliant and, as a result, it has a high elastic modulus, so $\theta$ was much lower than the one found for living animals [23], which have also a very high effective adhesion energy because they use a hierarchical peeling system and a smart body adaptation to optimize adhesion and detachment. Therefore, they can better adhere to surfaces yet detach quickly using low forces. Moreover, there are several other differences between adhesion of tapes and adhesion of biological systems: apart from the length scales, differences can be found in rate-dependency, reversibility, frictional behavior, in the bending resistance of the structure, etc..

## 4  Conclusions

In this paper, double peeling tests of standard adhesive tapes were performed with the aim of an *ad hoc* built experimental platform. Results show the existence of a critical peeling force above which complete detachment of the entire tape occurs. Moreover, during the process of detachment, the peeling angles move towards a limiting value, in correspondence of the maximum pull-off force.

The experimental data are in good agreement with theoretical calculations. In fact, the proposed theoretical model predicts the existence of a lower limit $\theta_{lim}$ of the peeling angle at which the pull-off force $F$ takes its maximum value $F_c$. Stable solutions below this threshold value are not possible.

Our experimental tests can be important for better understanding biological adhesive systems or anchorages, as well as for the design of bio-inspired super-adhesive smart materials or super-strong anchors. Moreover, our results could be useful for many applications, such as to design innovative adhesive systems, to determine new anchorage methods or to develop smart biomedical patches (e.g. self-adhering bandages or dressings).

**Acknowledgments**

N.M.P. is supported by the European Research Council PoC 2015 "Silkene" No. 693670, by the European Commission H2020 under the Graphene Flagship Core 1 No. 696656 (WP14 "Polymer Nanocomposites") and under the FET Proactive "Neurofibres" No. 732344. D.M. thanks financial support from the ERC Advanced Grant "Instabilities and nonlocal multiscale modelling of materials" FP7-PEOPLE-IDEAS-ERC-2013-AdG (2014-2019). L.A. and G.C. acknowledge Regione Apulia and the Italian Ministry of Education, University and Research for having supported the research activity within the project TRASFORMA Laboratory Network cod. 28.

**Conflict of Interest**

The authors declare that they have no conflict of interest.

This manuscript has not been published elsewhere and that it has not been submitted simultaneously for publication elsewhere.